\documentclass[letterpaper,USenglish]{oasics-v2018}

\usepackage{microtype}
\usepackage[inline]{enumitem}
\usepackage{comment}
\usepackage[super]{nth}
\usepackage{braket}

\title{QDB: From Quantum Algorithms Towards Correct Quantum Programs}

\author{Yipeng Huang}{Department of Computer Science, Princeton University. Princeton, NJ, USA.}{yipeng@cs.princeton.edu}{https://orcid.org/0000-0003-3171-6901}{This work is funded in part by EPiQC, an NSF Expedition in Computing,
under grant 1730082}

\author{Margaret Martonosi}{Department of Computer Science, Princeton University. Princeton, NJ, USA.}{mrm@princeton.edu}{}{}

\authorrunning{Y. Huang and M. Martonosi}

\Copyright{Yipeng Huang and Margaret Martonosi}

\subjclass{
}


\category{}

\relatedversion{}

\supplement{}

\funding{}

\acknowledgements{}

\EventEditors{Titus Barik, Joshua Sunshine, and Sarah Chasins}
\EventNoEds{3}
\EventLongTitle{9th Workshop on Evaluation and Usability of Programming Languages and Tools (PLATEAU 2018)}
\EventShortTitle{PLATEAU 2018}
\EventAcronym{PLATEAU}
\EventYear{2018}
\EventDate{November 5, 2018}
\EventLocation{Boston, Massachusetts, USA}
\EventLogo{}
\SeriesVolume{67}
\ArticleNo{4}
\nolinenumbers 

\begin{document}

\maketitle

\begin{abstract}
With the advent of small-scale prototype quantum computers, researchers can now code and run quantum algorithms that were previously proposed but not fully implemented.
In support of this growing interest in quantum computing experimentation, programmers need new tools and techniques to write and debug QC code.
In this work, we implement a range of QC algorithms and programs in order to discover what types of bugs occur and what defenses against those bugs are possible in QC programs.
We conduct our study by running small-sized QC programs in QC simulators in order to replicate published results in QC implementations.
Where possible, we cross-validate results from programs written in different QC languages for the same problems and inputs.
Drawing on this experience, we provide a taxonomy for QC bugs,
and we propose QC language features that would aid in writing correct code.

\end{abstract}

\section{Introduction}

Quantum computing is reaching an inflection point.
After years of work on both QC algorithms and low-level QC devices, small but viable QC prototypes are now available to run programs.
These QC prototypes are increasing in size,
with much research attention being placed on improving their reliability and increasing the counts of qubits (quantum bits),
the fundamental building block for QC~\cite{xrds, experimental_comparison, Preskill2018computinginnisqera}.

With small-scale machines available to run real code,
a natural challenge lies in creating correct and useful programs to run on them~\cite{nature, software_methodology}.
Until recently, QC algorithms were rarely programmed for actual execution, and therefore relatively little QC debugging has ever occurred.
Furthermore, QC debugging faces challenges beyond that of classical computing.
In particular, typical debugging approaches based on printing out variable values during program execution do not easily apply to QC programs,
because program states in QC ``collapse'' to classical values when observed.
Second, QC's ``no cloning rule'' precludes us from making a spare copy of variables to observe them elsewhere.
Third, while we have more freedom to observe states in QC simulations on classical computers, the massive state spaces of QC executions limits this approach to small programs.
Finally, even when limited simulations are tractable, it can be difficult to interpret the simulation results.

This paper surveys a range of QC algorithms and programs and offers a set of empirical and experiential insights on today's state-of-the-art in QC debugging.
For three benchmarks representing different application areas, we perform detailed debugging based on small-scale simulations.
For each, we give case studies of the types of bugs we found.
Most importantly, we use these experiences to assemble a set of ``design patterns for QC programming'' and related best practices in QC debugging.

In particular, the contributions of this paper are as follows:
\begin{itemize}
\item We specifically explore three major areas: quantum chemistry, integer factorization, and database search.
This is a broad spectrum of QC algorithms across not just application domains, but also problem size and algorithm strategies.
This allows us to point out particular domain-specific challenges or opportunities.
\item Where available, we study the same algorithm implemented in different languages or infrastructures.
From this, we draw comparative insights regarding how programming language or environment support can be useful in QC programming and debugging.
\item From these insights and experiences, we lay out a plan for debugging support in QC programming environments  to aid users in creating quantum code.
These include assertions, unit testing, code reuse, polymorphism, and QC-specific language types and syntax.
\end{itemize}

Overall, while QC programming has received significant prior attention and QC debugging has received some as well, our work offers steps forward in its detailed and comparative assessment across problem types and languages.
We see our work offering useful insights for QC programmers themselves, as well as language and system designers interested in building next-generation compilers and debuggers.
\section{Background on QC programming}
First, we review the principles of quantum computing~\cite{Kaye:2007:IQC:1206629, mermin2007quantum, metodi, nielsen_chuang},
in order to understand how writing correct quantum programs is different from classical programming.

\subsection{Qubits, superpositions, and entanglement}
 
The basic unit of information in QC is the qubit,
which can take on values of $\ket{0}$ and $\ket{1}$ like bits in classical computing,
but can also be viewed as a probabilistic ``superposition'' between the two values.
Quantum computers can also ``measure'' the value of a qubit,
forcing it to collapse out of superposition into a classical value such as `0' or `1'.
Measurement disturbs the values of variables in a quantum computer,
so we cannot easily pause execution and observe the values of qubits as a quantum program runs.




The state of individual qubits can be ``entangled'' together.
For this reason, as more qubits come into play in a quantum computer,
the number of states that data can be in grows exponentially.
For example, a two-qubit system can take on the values $\ket{00}, \ket{01}, \ket{10}, \ket{11}$, along with superpositions among these values;
furthermore, the two qubits can even be in a state of entanglement where the two cannot be treated as independent pieces of information.
A three qubit system has potential superpositions of eight states, and so on.
This exponential growth of possible values underlies the power of QC.

As a result of this large number of possible states, running a quantum program in simulation on a classical computer is costly.
Naive simulation of a 20-qubit quantum computer, for example, needs $2^{20}$ or roughly one million floating point numbers just to store the program state at any instant.
For this reason, testing and debugging quantum programs in simulation is only possible for toy-sized programs.


\subsection{Quantum computer operations, programs, and a taxonomy for bugs}

\begin{figure}
\centering
\includegraphics[width=0.6\columnwidth]{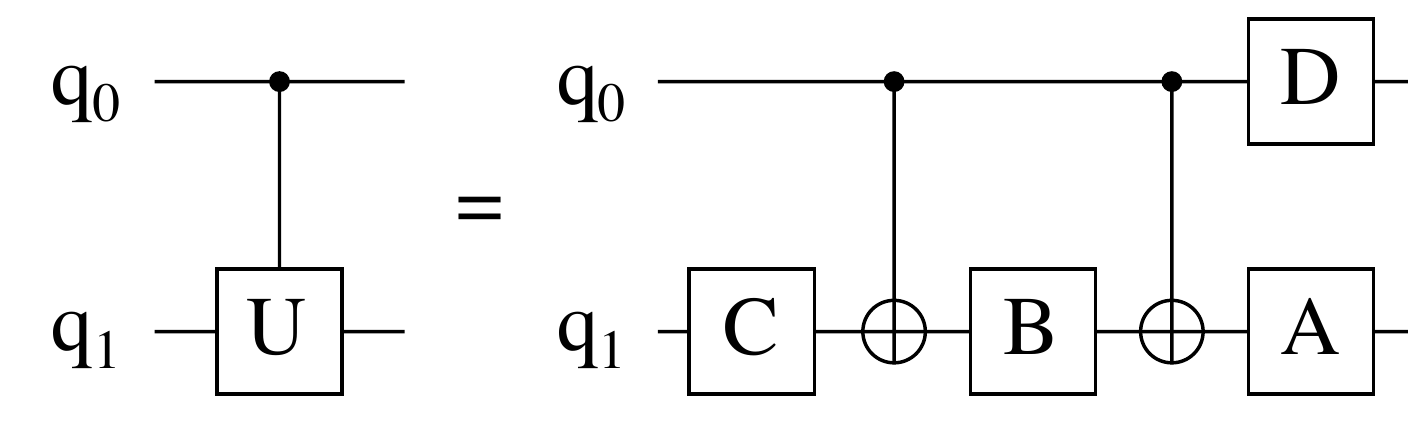}
\caption{
Decomposition of a simple QC program.
Time flows left to right, showing sequences of operations applied to qubits $q_0$ and $q_1$.
The left program is a ``controlled'' arbitrary operation $U$, which means whether the operation $U$ works on $q_1$ is dependent on the value of $q_0$.
The left sequence decomposes into the equivalent right sequence of more basic operations.
The basic operations include single-qubit ``rotations'' $A$ through $D$ that alter the probability distribution of qubit values.
The operations also include two two-qubit ``CNOT'' operations that flip a qubit (denoted $\oplus$) contingent on the value of another qubit (denoted $\bullet$)~\cite{nielsen_chuang}.}
\label{fig:decomposition}
\end{figure}

The process of quantum computing involves applying operations on qubits.
We use diagrams such as Figure~\ref{fig:decomposition} to represent sequences of quantum operations.
Looking at Figure~\ref{fig:decomposition} we see that quantum programs consist of three conceptual parts~\cite{hoare}:
\begin{enumerate}
\item \textbf{Inputs} to quantum algorithms include \emph{classical} input parameters such as coefficients for rotations $A$ through $D$, and \emph{quantum} initial values for qubits such as $q_0$ and $q_1$.
\item \textbf{Operations}, such as the specification of how a complex operation such as controlled arbitrary operation $U$ (Figure~\ref{fig:decomposition}, left) decomposes into basic operations $A$ through $D$ and CNOTs (Figure~\ref{fig:decomposition}, right).
Additionally, both basic and complex operations can be further composed according to patterns such as iteration, recursion, and mirroring.
\item \textbf{Outputs} of quantum algorithms are the final classical measurement values of qubits such as $q_0$ and $q_1$.
Furthermore, any temporary variables used in the course of a program have to be safely disentangled from the rest of the quantum state and discarded.
\end{enumerate}



Bugs in quantum programs can crop up due to mistakes made in any of these three parts of a QC program.
We will give examples of each kind of bug along with how to prevent them, using detailed case studies in the rest of this paper.
\subsection{QC algorithm primitives, benchmarks, and open source frameworks}
\label{sec:algorithms}

Given the rapid growth of QC infrastructure,
we now have a chance to test a variety of quantum algorithms written in many languages~\cite{comparison}.
Many different quantum algorithms rely on a handful of QC algorithm primitives to get speedups relative to classical algorithms~\cite{lanl_implementations, montanaro2016quantum, mosca2009quantum}.
Table~\ref{tab:benchmarks} classifies canonical quantum algorithms according to their algorithm primitives, and cites example implementations in different QC languages and tool chains.

This paper specifically focuses on program bugs and defenses in three areas: a quantum chemistry problem that uses quantum phase estimation, integer factorization using Shor's order finding algorithm, and Grover's database search algorithm.


Using programs written in the Scaffold language as a starting point~\cite{scaffcc}, we compile Scaffold code to OpenQASM, a QC assembly language~\cite{open_qasm}.
Then, we simulate the programs operation-by-operation in the QX simulator~\cite{qx}, in order to see their intermediate states and outputs.
We cross reference the programs' results against implementations in other languages, such as LIQUi|>~\cite{design_automation}, ProjectQ~\cite{emulation, Steiger2018projectqopensource} and Q\#~\cite{q_sharp}.
From this debugging experience we identify possible bugs and defenses.
Furthermore, we review the codes across languages to understand the relative merits of different QC language features.


\begin{table*}[t]
\caption{Quantum algorithm primitives and open source benchmarks in open source tool chains.}
\small
\centering
\begin{tabularx}{\linewidth}
{ |p{1.85cm}|p{3.5cm}|X| }
\hline
\textbf{Primitives} & \textbf{Quantum algorithms} & \textbf{Benchmark implementations} \\
\hline
\hline

\multirow{2}{1.85cm}{Entanglement protocols}
& \multirow{2}{3.5cm}{superdense coding / quantum teleportation}
& Q\# teleportation~\cite{q_sharp} \\
& & pyQuil teleportation~\cite{quil} \\

\hline

\multirow{3}{1.85cm}{Quantum (random) walks}
& tree traversal
& Scaffold / Quipper binary welded tree~\cite{quipper, scaffcc, quipper_cacm} \\
\cline{2-3}
& graph traversal
& Scaffold / Quipper triangle finding problem~\cite{quipper, scaffcc, quipper_cacm} \\
\cline{2-3}
& satisfiability
& Scaffold / Quipper Boolean formula~\cite{quipper, scaffcc, quipper_cacm} \\
\hline

\multirow{3}{*}{Adiabatic}
& Ising spin model
& Scaffold / Q\# adiabatic Ising model~\cite{scaffcc, q_sharp} \\
\cline{2-3}
& \multirow{2}{3.5cm}{quantum approximate optimization algorithm}
& QISKit Aqua QAOA \\
& & pyQuil QAOA ansatz~\cite{quil} \\

\hline

\multirow{3}{1.85cm}{Variational Quantum Eigensolver}
& \multirow{3}{*}{Hamiltonian simulation}
& QISKit Aqua quantum chemistry \\
& & Q\# $\mathrm{H_2}$ simulation~\cite{q_sharp} \\
& & Rigetti Grove VQE~\cite{quil} \\

\hline


\multirow{5}{1.85cm}{Quantum Fourier Transform (QFT)}
& phase estimation
& Scaffold / Quipper ground state estimation~\cite{quipper, scaffcc, quipper_cacm} \\
\cline{2-3}
& period finding & Scaffold class number~\cite{scaffcc} \\
\cline{2-3}
& order finding & Scaffold / ProjectQ / Q\# Shor's factoring~\cite{scaffcc, Steiger2018projectqopensource, q_sharp} \\
\cline{2-3}
& hidden subgroup problem & Quipper unique shortest vector~\cite{quipper, quipper_cacm} \\
\cline{2-3}
& linear algebra & Quipper quantum linear systems~\cite{quipper, quipper_cacm} \\

\hline

\multirow{2}{1.85cm}{Amplitude amplification}
& \multirow{2}{*}{database search}
& Scaffold square root~\cite{scaffcc} \\
& & ProjectQ / Q\# Grover's database search~\cite{Steiger2018projectqopensource, q_sharp} \\

\hline

\end{tabularx}
\label{tab:benchmarks}
\end{table*}



\section{Case study: Quantum chemistry}
First, we discuss our experience building up and debugging a simple quantum chemistry program.
Quantum chemistry problems entail finding properties of molecules from theoretical first principles~\cite{qchem, qchem_nsf}.
Researchers anticipate these will be the first applications for QC due to the relatively few number of qubits they need to surpass classical computer algorithms.
Debugging these problems is distinctively challenging, due to the importance of getting a large number of classical input parameters all correct, and because of the dearth of physically meaningful intermediate states we can check in the course of algorithm execution.

\subsection{Bug type 1: Incorrect classical input parameters}
\label{sec:classical_bug}
A key part of quantum chemistry programs is in correctly building up a ``Hamiltonian'' subroutine that simulates inter-electron forces.
The procedure for doing this was laid out in detail by Whitfield~\cite{whitfield}.
We followed this procedure to create a subroutine for simulating the hydrogen molecule, but we needed additional validation from several other sources to get a bug-free subroutine~\cite{gate_count}.
These resources include raw chemistry data found in open source repositories for the LIQUi|> framework\footnote{\url{https://github.com/StationQ/Liquid/blob/master/Samples/h2_sto3g_4.dat}
}.
The final parameters for actual operations on qubits were validated against a follow-up paper~\cite{bk_transform} and an implementation in the QISKit framework\footnote{\url{https://github.com/Qiskit/aqua/blob/master/test/H2-0.735.json}}.
Because the procedure for preparing these quantum chemistry models involves many steps and needs domain expertise, software packages such as OpenFermion now automate this process~\cite{openfermion}.
Nonetheless, there is room for improvement in standardizing input data formats to eliminate bugs in this process.

Once the Hamiltonian subroutine is built, we can use the model in a variety of quantum algorithms spanning different primitives in Table~\ref{tab:benchmarks}.
These include phase estimation (an application of quantum Fourier transforms)~\cite{qpe}, variational quantum eigensolvers~\cite{vqe}, and adiabatic algorithms~\cite{adiabatic}.
In this paper, we use iterative phase estimation to find the ground state energy of our $\mathrm{H_2}$ model, validating results published by Lanyon~\cite{lanyon}.

\begin{table}[t]
\small
\centering
\caption{QC calculated energy for $\mathrm{H_2}$ (bond length = 73.48 pm) for different electron assignments.}
\label{fig:hydrogen}
\begin{tabular}{|r||p{.75cm}|p{.75cm}|p{.75cm}|p{.75cm}||l|}
\hline
&\multicolumn{4}{c||}{\textbf{Electron assignments}}&\multirow{3}{3cm}{\textbf{QC calculated energy (relative)}}\\
&\multicolumn{2}{c|}{Bonding}&\multicolumn{2}{c||}{Antibonding}&\\
&$\uparrow$&$\downarrow$&$\uparrow$&$\downarrow$&\\
\hline
\nth{3} excited state (E3)&0&0&1&1&-0.164\\
\hline
\multirow{2}{*}{\nth{2} excited state (E2)} &0&1&1&0&\multirow{2}{*}{-0.217}\\
&1&0&0&1&\\
\hline
\multirow{2}{*}{\nth{1} excited state (E1)} &0&1&0&1&\multirow{2}{*}{-0.244}\\
&1&0&1&0&\\
\hline
Ground state (G)&1&1&0&0&-0.295\\
\hline
\end{tabular}
\label{tab:hydrogen}
\end{table}

\subsection{Bug type 2: Incorrect quantum initial values}
\label{sec:allocation}

The correct preparation of qubit initial values is important.
Incorrect initial values would cause the program to find solutions to different problems altogether.
In this quantum chemistry problem, the initial values control the locations of the two electrons in $\mathrm{H_2}$.
As shown in Table~\ref{tab:hydrogen}, we need the qubit assignment for finding the ground energy of $\mathrm{H_2}$, while other assignments lead to results for other energy levels.

The symmetry of $\mathrm{H_2}$ allows us to perform a sanity check, to make sure the Hamiltonian and the iterative phase estimation subroutines are working correctly.
Though there are six ways to assign two electrons to four locations,
there are in fact only four distinct energy levels, as shown in the experimental data.
Checking that the two different ways to obtain E1 (and E2) give the same energy levels validates that the model correctly preserves symmetry.

\subsection{Defense type 1: Assertions on algorithm preconditions}
\label{sec:preconditions}

Given how important correct initial values are for all quantum algorithms, it is worthwhile to explicitly check for these algorithm preconditions before continuing with execution or simulation.
What the preconditions should be depends on the type of algorithm.
For example, the phase estimation subroutine in this case study (along with other algorithms relying on quantum Fourier transforms), expect inputs that are maximally in superposition among all possible values.
Likewise, ``ancillary qubits'' such as the inputs to the Hamiltonian subroutine take on completely classical (integer) initial values.
Lastly, quantum protocols often need to start with entangled states.
These required input states are among the few places in quantum algorithms where we can check states for specific values.
We can check these preconditions by running or simulating programs up to the entry point of subroutines, and performing a premature measurement to check for these anticipated states, finally restarting the program knowing that execution is correct up to that point.
Thus far, the Q\# framework has the most extensive support for precondition checking~\cite{q_sharp}.


\subsection{Defense type 2: Assertions on algorithm progress}
\label{sec:progress}

Unlike the other two case studies later in this paper, the debugging process for the quantum chemistry benchmark is coarse-grained.
That is because the Hamiltonian subroutine is a monolithic block of code whose components do not have obvious expected outputs---its components represent pair-wise electron interactions, and do not have inherent physical meaning.
So how do we debug this program? The preconditions in the last section make sure the inputs to the algorithm are correct; the other observable state we have for debugging is to check the behavior of the algorithm as a whole.

In this quantum chemistry program, we can check for two types of overall algorithm behavior.
One is the solution should converge to a steady value as finer Trotter time steps (a kind of numerical approximation) are chosen; a lack of this type of convergence indicates a bug in the Hamiltonian subroutine.
The other algorithm behavior is when we vary the precision of the phase estimation algorithm,
the most significant bits of the measurement output sequences should be the same---in other words, rounding the output of a high-precision experiment should yield the same output as a lower-precision experiment.
a lack of this convergence indicates a bug in the iterative phase estimation subroutine.
These checks for expected algorithm progress also apply to other algorithms.



\section{Case study: Shor's algorithm for integer factorization}

\begin{table*}
\footnotesize
\centering
\caption{Shor's factorization algorithm subroutines~\cite[p.~25]{metodi}.}
\begin{tabularx}{\linewidth}{ |X|X| }
\hline
\textbf{Program subroutine code} & \textbf{Shared library code} \\
\hline
\hline
\parbox{\columnwidth}{
Shor's routine for factoring 15;\\
calculating powers of a number
\begin{itemize}
    \item controlled modular multiplication
    \item controlled modular addition
    \item controlled addition
\end{itemize}
} & \parbox{\columnwidth}{
\begin{itemize}
    \item quantum Fourier transform
    \item controlled controlled rotation
    \item controlled rotation
    \item controlled swap
    \item swap
\end{itemize}
}\\\hline
\end{tabularx}
\label{tab:shor}
\end{table*}

While our debugging strategy for quantum chemistry had to be coarse-grained, the debugging process for Shor's algorithm in this section allows us to look inside the program one subroutine at a time, where we can compare the intermediate results against known expected values.

Shor's factorization algorithm uses a quantum computer to factor a composite number in polynomial time complexity, providing exponential speedup relative to the best known classical algorithms~\cite{shor}.
We follow an example for an implementation that minimizes the qubit cost~\cite{beauregard}, and replicate results for factoring 15, the simplest example~\cite{experimental_demo_shor}~\cite[p. 235]{nielsen_chuang}.

\subsection{Bug type 3: Incorrect operations and transformations}
\label{sec:basic}

\begin{table*}[t]
\caption{
Correct and incorrect code for rotation decomposition.
Using the Scaffold language~\cite{scaffcc} as an example,
we code out the controlled operation U in Figure~\ref{fig:decomposition} where U is a rotation in just one axis.
Because only one axis is needed, we can drop either operation A or C, paying attention to the sign on the angles.
Reordering the lines of code or signs results in a rotation in the wrong direction.
}
\small
\centering
\begin{tabularx}{\textwidth}{|X|X||p{4cm}|}
\hline
{Correct, operation A unneeded}
&
{Correct, operation C unneeded}
&
{Incorrect, angles flipped}\\
\hline
\texttt{Rz(q1,+angle/2); // C} &
\texttt{CNOT(q0,q1);} &
\texttt{Rz(q1,-angle/2);} \\
\texttt{CNOT(q0,q1);} &
\texttt{Rz(q1,-angle/2); // B} &
\texttt{CNOT(q0,q1);} \\
\texttt{Rz(q1,-angle/2); // B} &
\texttt{CNOT(q0,q1);} &
\texttt{Rz(q1,+angle/2);} \\
\texttt{CNOT(q0,q1);} &
\texttt{Rz(q1,+angle/2); // A} &
\texttt{CNOT(q0,q1);} \\
\texttt{Rz(q0,+angle/2); // D} &
\texttt{Rz(q0,+angle/2); // D} &
\texttt{Rz(q0,+angle/2); // D} \\
\hline\end{tabularx}
\label{tab:bug_example}
\end{table*}

In order to correctly implement Shor's algorithm we first have to build up the quantum subroutines shown in Table~\ref{tab:shor}.
These basic subroutines can be tricky to get right.
Take the controlled rotation in Figure~\ref{fig:decomposition} as an example:
Table~\ref{tab:bug_example} shows multiple ways to code the decomposition of the controlled rotation, and small mistakes can lead to incorrect behavior.


\subsection{Defense type 3: Language support for subroutines / unit tests}
\label{sec:modularization}

An obvious defense against coding mistakes in basic subroutines is to use a library of shared code.
Doing so helps ensure program correctness by allowing programmers to exhaustively validate small subroutines,
in order to bootstrap larger subroutines.
Unit testing is especially important in QC as running or simulating large quantum programs is impossible for now.

An additional benefit is logically structured code allows compilers to select the best concrete implementation for the abstract functionality the programmer needs, based on hardware constraints and input parameters~\cite{hoare}.
For example, the most cost-efficient implementation for modular exponentiation in Shor's factorization algorithm depends on how many qubits are available:
the compiler can choose from minimum-qubit~\cite{beauregard, toffoli_modular, Takahashi} or minimum-operation~\cite{min_depth} implementations for the arithmetic subroutines.



\subsection{Bug type 4: Incorrect composition of operations using iteration}
\label{sec:iterate}

\begin{figure*}
\lstset{
    language=C,
    numbers=right,
    caption={Controlled adder subroutine using Fourier transform in the Scaffold language~\cite{scaffcc}.},
    label=lst:scaffold_adder
}
\begin{lstlisting}
// outputs a + b, where a is a `width' bit constant integer
// b is an integer encoded on `width' qubits in Fourier space
module cADD (
  const unsigned int c_width, // number of control qubits
  qbit ctrl0, qbit ctrl1, // control qubits
  const unsigned int width, const unsigned int a, qbit b[]
) {
  for (int b_indx=width-1; b_indx>=0; b_indx--) {
    for (int a_indx=b_indx; a_indx>=0; a_indx--) {
      if ((a >> a_indx) & 1) { // shift out bits in constant a
        double angle = M_PI/pow(2,b_indx-a_indx); // rotation angle
        switch (c_width) {
          case 0: Rz ( b[b_indx], angle ); break;
          case 1: cRz ( ctrl0, b[b_indx], angle ); break;
          case 2: ccRz ( ctrl0, ctrl1, b[b_indx], angle ); break;
}}}}}
\end{lstlisting}
\end{figure*}

Once we have built our basic subroutines, a common pattern in quantum programs is to use iterations to compose subroutines.
Listing~\ref{lst:scaffold_adder} shows the iteration code for a constant-value adder, showing tricky places in lines 8 through 11 for bugs to crop up,
including indexing errors, bit shifting errors, endian confusion, and mistakes in rotation angles.
In general this type of iteration code is commonplace in programs that rely on quantum Fourier transforms.

\subsection{Defense type 4: Language support for numerical data types}
\label{sec:numerical}

One way to defend against bugs in iteration code is to introduce QC data types for numbers,
providing greater abstraction than working with raw qubits.
For example, ProjectQ has quantum integer data types~\cite{Steiger2018projectqopensource},
while Q\#~\cite{q_sharp} and Quipper~\cite{quipper, quipper_cacm} offer both big endian and little endian versions of subroutines involving iterations.
These QC data types permit useful operators (e.g., checking for equality) that help with debugging and writing assertions.




\begin{table*}
\small
\centering
\caption{Correct classical input $a$ and $a^{-1}$ to Shor's algorithm for factoring 15, using 7 as a guess.}
\begin{tabular}{ |r||p{1cm}|p{1cm}|p{1cm}|p{1cm}|p{1cm}| } 
\hline
\textbf{$k$, the algorithm iteration} & \textbf{0} & \textbf{1} & \textbf{2} & \textbf{3} & \textbf{\ldots} \\
\hline
$a=7^{2^k} \mod 15$ & 7 & 4 & 1 & 1 & \ldots \\
\hline
$a^{-1}$; $a\times a^{-1} \equiv 1 \mod 15$ & 13 & 4 & 1 & 1 & \ldots \\
\hline
\end{tabular}
\label{tab:shor_inputs}
\end{table*}

\subsection{Bug type 5: Incorrect deallocation of qubits}
\label{sec:deallocation}

Variable scoping is an important language feature in classical computing that ensures proper data encapsulation.
In QC, scoping is similarly important for temporary variables known as ``ancillary qubits.''
Anything that happens to a subroutine's ancillary qubits---such as measurement, reinitialization, or lapsing into decoherence---may have unintended effects on the subroutine's outputs\footnote{As an analogy in classical computing, it is as if accessing an out-of-scope variable can still affect program state; while such behavior is unintuitive, it is a result of how entanglement works in QC.}. 
Because improper ancillary qubit deallocation can lead to wrong results,
it is important for subroutines to reverse their operations on their ancillary qubits,
so that they properly undo any entanglement between the ancillary and output qubits.

We can demonstrate a bug involving incorrect qubit deallocation,
by deliberately making a mistake while reversing operations in a subroutine.
For example, Shor's algorithm relies on correct pairs modular inverse numbers as input parameters, such as those in Table~\ref{tab:shor_inputs}.
By feeding an incorrect pair of inputs (e.g., replacing 13 with a 12),
the algorithm proceeds to possibly give us wrong output values, as shown in Table~\ref{tab:shor_outputs}.
At the same time, the mistake prevents the modular multiplication operation from being properly reversed,
which has the effect of preventing the ancillary qubits from properly disentangling with other qubits,
so they fail to return to their initial values at the end of the algorithm.

\subsection{Defense type 5: Assertions on algorithm postconditions}
\label{sec:postconditions}

\begin{table*}
\footnotesize
\centering
\caption{Probability of measuring values of outputs and ancillary qubits of Shor's algorithm, with incorrect inputs ($a^{-1}=12$ instead of 13 on first iteration).
If the ancillary qubits collapse to zero on measurement, the algorithm still succeeds, returning correct outputs of 0, 2, 4, 6~\cite[p. 235]{nielsen_chuang}.
However, the possibility of measuring non-zero for the ancillary qubits indicates a bug.
}
\begin{tabular}{ |c r|p{.666cm}p{.666cm}p{.666cm}p{.666cm}p{.666cm}p{.666cm}p{.666cm}p{.666cm}| } 
\hline
\multicolumn{2}{|c|}{\multirow{2}{*}{\textbf{Probability}}} & \multicolumn{8}{c|}{\textbf{Output measurement}} \\
&& 0 & 1 & 2 & 3 & 4 & 5 & 6 & 7 \\
\hline
\multirow{3}{*}{\textbf{Ancillary}} & 0 & 1/8 & 0 & 1/8 & 0 & 1/8 & 0 & 1/8 & 0 \\
\multirow{3}{*}{\textbf{qubit}} & 2 & 1/64 & 1/64 & 1/64 & 1/64 & 1/64 & 1/64 & 1/64 & 1/64 \\
\multirow{3}{*}{\textbf{measurement}} & 7 & 1/64 & 1/64 & 1/64 & 1/64 & 1/64 & 1/64 & 1/64 & 1/64 \\
& 8 & 1/64 & 1/64 & 1/64 & 1/64 & 1/64 & 1/64 & 1/64 & 1/64 \\
& 13 & 1/64 & 1/64 & 1/64 & 1/64 & 1/64 & 1/64 & 1/64 & 1/64 \\
\hline
\end{tabular}
\label{tab:shor_outputs}
\end{table*}

We can use postconditions at the end of algorithms to detect bugs that lead to incorrect deallocation of ancillary qubits.
Continuing with our example in Table~\ref{tab:shor_outputs},
we see that the cases where ancillary qubits collapse to anything other than zero correspond to cases where the outputs are wrong.
That is because the ancillary qubits remain improperly entangled with the output qubits at the end of the algorithm.
We can detect these buggy outputs by asserting that ancillary qubits should always return to their initial values.
The significance of these observations is that when algorithms work correctly, we typically do not care to measure the value of ancillary qubits as they do not contain information.
But in buggy QC algorithm implementations, they are useful side channels for debugging.
\section{Case study: Grover's algorithm for database search}

\newcommand{\specialcell}[2][c]{%
  \begin{tabular}[#1]{@{}l@{}}#2\end{tabular}}

\begin{table*}
\caption{Grover's amplitude amplification subroutine in two languages, showcasing QC-specific language syntax for reversible computation (rows 2 \& 6) and controlled operations (rows 3 \& 5).}
\begin{tabular}{ |r|l|l| }
\hline
& \textbf{Scaffold (C syntax)~\cite{scaffcc}} & \textbf{ProjectQ (Python syntax)~\cite{Steiger2018projectqopensource}} \\\hline
\hline

\textbf{1}&

\specialcell{
\texttt{int j;}\\
\texttt{qbit ancilla[n-1]; // scratch register}\\
\texttt{for(j=0; j<n-1; j++) PrepZ(ancilla[j],0);}
}&

\specialcell{
\texttt{\# reflection across}\\
\texttt{\# uniform superposition}
}

\\\hline

\textbf{2}&

\specialcell{
\texttt{// Hadamard on q}\\
\texttt{for(j=0; j<n; j++) H(q[j]);}\\
\texttt{// Phase flip on q = 0...0 so invert q}\\
\texttt{for(j=0; j<n; j++) X(q[j]);}
}&

\specialcell{
\texttt{with Compute(eng):}\\
\texttt{~~~~All(H) | q}\\
\texttt{~~~~All(X) | q}
}

\\\hline

\textbf{3}&

\specialcell{
\texttt{// Compute x[n-2] = q[0] and ... and q[n-1]}\\
\texttt{CCNOT(q[1], q[0], ancilla[0]);}\\
\texttt{for(j=1; j<n-1; j++)}\\
\texttt{~~~~CCNOT(ancilla[j-1], q[j+1], ancilla[j]);}
}&

\specialcell{
\texttt{with Control(eng, q[0:-1]):}
}

\\\hline

\textbf{4}&

\specialcell{
\texttt{// Phase flip Z if q=00...0}\\
\texttt{cZ(ancilla[n-2], q[n-1]);}
}&

\specialcell{
\texttt{~~~~Z | q[-1]}
}

\\\hline

\textbf{5}&

\specialcell{
\texttt{// Undo the local registers}\\
\texttt{for(j=n-2; j>0; j--)}\\
\texttt{~~~~CCNOT(ancilla[j-1], q[j+1], ancilla[j]);}\\
\texttt{CCNOT(q[1], q[0], ancilla[0]);}
}&

\specialcell{
\texttt{\# ProjectQ automatically}\\
\texttt{\# uncomputes control}
}

\\\hline

\textbf{6}&

\specialcell{
\texttt{// Restore q}\\
\texttt{for(j=0; j<n; j++) X(q[j]);}\\
\texttt{for(j=0; j<n; j++) H(q[j]);}
}&

\specialcell{
\texttt{Uncompute(eng)}
}

\\\hline
\end{tabular}
\label{fig:grover_code}
\end{table*}

So far, we have presented defenses against bugs following two general strategies.
One is to use assertions to detect when and where the program has a bug.
The other is to use quantum-specific programming language features to prevent bugs altogether: these features include support for subroutines and numerical types for quantum data.
Here in this section, we use the Grover's benchmark to showcase two more language features for common QC program patterns: reversible computation and controlled operations.

Grover's search algorithm finds an entry that matches search criteria, among an input data set of size $N$, with a time cost on the order of $\sqrt{N}$.
That represents a polynomial speedup relative to the linear time cost in a classical computer~\cite{grover2001schrodinger}.

The Grover's algorithm comprises three parts.
First, the input qubits representing the indices of the matching entries are put in a state of superposition, akin to querying all entries at once.
Second, the queries are put through a subroutine that checks for the search criteria.
In this case study, our criteria is to find the square root of a number in a Galois field of two elements, a simple abstract algebra setting.
Finally in the critical step, the amplitude amplification algorithm primitive amplifies the index that matches the criteria while damping out those that do not.
The operations in this final step are prime examples of two QC program patterns, reversible computation and controlled operations.
We show in Table~\ref{fig:grover_code} how these code patterns are written in two languages, Scaffold~\cite{scaffcc} and ProjectQ~\cite{Steiger2018projectqopensource}.


\subsection{Bug type 6: Incorrect composition of operations using mirroring}
\label{sec:mirroring}
Section~\ref{sec:deallocation} discussed how bugs in deallocating ancillary qubits can happen due to bad parameters.
Here we see how bugs in deallocating ancillary qubits can happen due to incorrect composition of operations following a mirroring pattern.
For example, in Table~\ref{fig:grover_code}, the operations in rows 2 and 3 are respectively mirrored and undone in rows 6 and 5.
These lines of code need careful reversal of every loop and every operation.

\subsection{Defense type 6: Language support for reversible computation}
\label{sec:reversible}
Syntax support for reversible computation,
such as that in ProjectQ~\cite{Steiger2018projectqopensource}, automatically mirrors and inverts sequences of operations,
shortening code and reducing mistakes.

\subsection{Bug type 7: Incorrect composition of operations using recursion}
\label{sec:recursion}
A common pattern in quantum programs involves performing operations (e.g., add), contingent on a set of qubits known as control qubits. 
Without language support, this pattern needs many lines of code and manual allocation of ancillary qubits.
In the Scaffold code example in Table~\ref{fig:grover_code},
rows 3 and 5 are just computing the intersection of qubits \texttt{q},
with the help of ancillary qubits initialized in row 1,
in order to realize the controlled rotation operation in row 4.
Furthermore, quantum algorithms often need varying numbers of control qubits in different parts of the algorithm, leading to replicated code from multiple versions of the same subroutine differing only by the number of control qubits\footnote{
An example appeared in the Shor's case study Listing~\ref{lst:scaffold_adder}.
The addition operation was contingent on control qubits taken as parameters in lines 4 and 5.
Depending on how many control qubits were needed,
the switch statement in lines 12 through 15 applied the correct operation.}.

\subsection{Defense type 7: Language support for controlled operations}
\label{sec:controlled}
Language support for controlled operations (e.g,  ProjectQ) shortens code, preventing mistakes.


\section{Conclusion}

\begin{table*}
\small
\centering
\caption{Applicability of defense strategies (down) against location of QC program bugs (across).}
\begin{tabular}{ |l|l||c|c|c|c|c|c|c| }
\hline
\multicolumn{2}{|l||}{\multirow{3}{*}{\textbf{}}} & \multicolumn{2}{c|}{input} & \multicolumn{4}{c|}{operations} & output \\\cline{3-9}
\multicolumn{2}{|l||}{} & classical & qubit & \multirow{2}{*}{basic} & \multirow{2}{*}{iterate} & \multirow{2}{*}{mirror} & \multirow{2}{*}{recurse} & qubit \\
\multicolumn{2}{|l||}{} & params. & alloc. & \multirow{2}{*}{\S\ref{sec:basic}} & \multirow{2}{*}{\S\ref{sec:iterate}} & \multirow{2}{*}{\S\ref{sec:mirroring}} & \multirow{2}{*}{\S\ref{sec:recursion}} & dealloc. \\
\multicolumn{2}{|l||}{} & \S\ref{sec:classical_bug} & \S\ref{sec:allocation} & & & & & \S\ref{sec:deallocation} \\
\hline
\hline
\multirow{4}{1.15cm}{QC specific lang. features} & unit testing \S\ref{sec:modularization} & & \checkmark & \checkmark & \checkmark & \checkmark & \checkmark & \checkmark \\\cline{2-9}
& data types \S\ref{sec:numerical} & & & & \checkmark & & & \\\cline{2-9}
& reverse comp. \S\ref{sec:reversible} & & & & \checkmark & \checkmark & & \checkmark \\\cline{2-9}
& controlled ops. \S\ref{sec:controlled} & & & & \checkmark & \checkmark & \checkmark & \\\hline
\multirow{3}{1.15cm}{Assertion checks} & preconditions \S\ref{sec:preconditions} & & \checkmark & & & & & \\\cline{2-9}
& algo progress \S\ref{sec:progress} & \checkmark & \checkmark & \checkmark & \checkmark & \checkmark & \checkmark & \checkmark \\\cline{2-9}
& postconds. \S\ref{sec:postconditions} & \checkmark & \checkmark & \checkmark & \checkmark & \checkmark & \checkmark & \checkmark \\\hline
\end{tabular}
\label{tab:apply_matrix}
\end{table*}


For the first time, we have access to comprehensive and representative program benchmarks for all major areas of quantum algorithms, implemented in multiple languages, along with input datasets and outputs that are detailed enough to permit cross-validation.
Using our experience running and debugging these programs, we presented in this paper defense strategies that facilitate writing bug-free QC code, summarized in Table~\ref{tab:apply_matrix}.
Successful transplantation of these ideas from classical languages to QC languages can pave the way towards correct and useful quantum programs.

\appendix


\clearpage
\bibliography{qdb_plateau}
\end{document}